\documentclass[runningheads]{llncs}
\usepackage[T1]{fontenc}
\usepackage{cite}
\usepackage{amsmath,amssymb,amsfonts}
\usepackage{algorithmic}
\usepackage{graphicx}
\usepackage{textcomp}
\usepackage{xcolor}
\usepackage{multirow}
\usepackage{tabularx}
\usepackage{booktabs}
\usepackage{array}
\usepackage{float}

\begin{document}

\newcommand{\AR}[1]{\textcolor{blue}{#1}}
\def\BibTeX{{\rm B\kern-.05em{\sc i\kern-.025em b}\kern-.08em
    T\kern-.1667em\lower.7ex\hbox{E}\kern-.125emX}}
\title{Assessing Generalisation Capability of Machine Learning Models for Intrusion Detection}
\titlerunning{Generalisation Capability of ML for Intrusion Detection}
%
\author{Md Zakir Hossain\inst{1}\orcidID{0000-0003-1892-831X} \and
Md Ayshik Rahman Khan\inst{2}\orcidID{0000-0003-0225-6969} 
\and
Md Rafiqul Islam\inst{3}\orcidID{0000-0001-8317-5727} 
\and
Syed Mohammed Shamsul Islam\inst{4}\orcidID{0000-0002-3200-2903}
\and
Tom Gedeon\inst{5}\orcidID{0000-0001-8356-4909}}
\authorrunning{MZ. Hossain et al.}
%
\institute{The Australian National University, Canberra ACT 2600 Australia 
\and
The University of Melbourne, Parkville VIC 3010 Australia
\and
Charles Sturt University, Albury-Wodonga, NSW 2640 Australia
\and
Edith Cowan University, Joondalup WA 6027 Australia
\and
Curtin University, Perth WA 6102 Australia\\
\email{zakir.hossain@anu.edu.au}}
\maketitle              
\begin{abstract}
The growth of networked and IoT systems has intensified cyber-security threats and exposed the limits of traditional signature-based intrusion detection. Although machine-learning-based intrusion detection systems often report strong benchmark performance, high accuracy within a single dataset does not necessarily guarantee reliable performance in unseen network environments. This study investigates the generalisation capability of supervised machine learning models for intrusion detection using UNSW-NB15 and TON\_IoT. Random Forest, Logistic Regression, and Naive Bayes were evaluated under same-dataset and cross-dataset settings. Random Forest achieved the strongest same-dataset performance, with 95.08\% accuracy on UNSW-NB15 and 99.79\% on TON\_IoT, but performance dropped sharply in cross-dataset testing. When trained on UNSW-NB15 and tested on TON\_IoT or vice versa, below 40\% accuracy. These results reveal a significant generalisation gap in intrusion detection. We connect this challenge to affective computing and human-centric AI, where behavioural signal analysis, anomaly detection, domain shift, and context-sensitive modelling are also central. This framing highlights the need for adaptive, generalisable cyber-security models that can operate across changing network and IoT environments.

\end{abstract}
\section{Introduction}

The widespread adoption of computer networks and digital services has made cybersecurity a critical concern in modern information systems \cite{saranya2020performance}. Along with the continued expansion of network infrastructures, cyberattacks have also become increasingly frequent, diverse, and sophisticated, targeting both individual users and large-scale organizations. Attacks such as distributed denial-of-service (DDoS) attacks, port scanning, brute-force attacks, and web-based exploits have become extremely common\cite{lu2025adversarial}. According to a report by IBM, 97\% organizations reported an AI based security issue, and the global average cost of data breaches was 4.4 million USD \cite{IBM}. These data breaches pose a challenge to existing security mechanisms and necessitate the development of more intelligent intrusion detection systems (IDS).

Intrusion Detection Systems (IDS) are commonly used to identify abnormal or malicious activities within network traffic. Traditional intrusion detection approaches generally follow two main techniques: signature-based detection and anomaly-based detection \cite{axelsson2000intrusion}. However,  traditional methods of intrusion detection often require manual detection, and these practices are labor-intensive and increasingly inadequate for identifying modern and evolving cyberattacks \cite{aljanabi2021intrusion}. Also, many existing IDS solutions rely on outdated datasets such as KDDCup99, which was built back in 1999 and lacks data on contemporary threats such as advanced DDOS, XSS or SQL injection \cite{kenyon2020public} \cite{adek2020survey}, and according to G{\"u}m{\"u}{\c{s}}ba{\c{s}} et al. \cite{gumucsbacs2020comprehensive}, the performance of the model strongly depends on the quantity and quality of the data set available. These limitations leave networks vulnerable and motivate the development of more effective detection mechanisms.

To overcome these limitations, machine learning has been increasingly applied to intrusion detection. Prior research has explored a range of machine learning techniques for detecting and classifying cyberattacks, demonstrating that data-driven models can improve detection performance compared to traditional approaches \cite{khan2024ai} \cite{li2017intrusion}. However, challenges such as false positives, data quality, adversarial inputs, and implementation cost limit reliability and so are still a concern, despite their effectiveness against evolving and zero-day attacks.

Beyond these general challenges, recent studies have found that model performance in intrusion detection is also influenced by dataset characteristics and evaluation design. In particular, it was noted that dataset diversity and class imbalance can create biases in performance, and that preprocessing and clustering choices can significantly influence detection outcomes \cite{gumucsbacs2020comprehensive}. This makes it difficult to interpret model performance unless evaluation is carried out carefully and consistently across attack categories. In this context, \cite{maseer2021benchmarking} compared a wide range of supervised and unsupervised methods and highlighted class imbalance as a major challenge. They emphasized multi-criteria evaluation when assessing IDS models, rather than relying on a single metric. Similarly, Panwar et. al \cite{panwar2022intrusion} described an IDS workflow using CICIDS2017 with preprocessing, feature selection, and classification, and reported Random Forest classifiers as achieving the highest accuracy across binary and multiclass classification.

Many studies that explored comparative results across datasets and model families also suggest that no single method performs best for all attack types. Almseidin et al. \cite{almseidin2017evaluation} evaluated multiple learning algorithms on KDD-style data and concluded that a single machine learning algorithm cannot handle all categories of attacks efficiently. In more specialized scenarios, very high performance has been reported for targeted detection tasks—for example, botnet detection using the CSE-CIC-IDS2018 dataset with high accuracy and a low false positive rate \cite{kanimozhi2019calibration}. Taken together, these studies reinforce the view that effectiveness of intrusion detection systems depends on both the dataset and the attack context, and cross-dataset evaluation still remains limited, which makes generalization to real-world scenarios harder to assess.

Although many studies report high accuracy on individual benchmark datasets, evaluation within the same-dataset does not fully show whether a model can perform well in a different network environment. Intrusion detection datasets can differ in traffic generation methods, attack patterns, feature distributions, and data collection environments. As a result, a model may perform well when trained and tested on the same dataset but fail to generalize when applied to another dataset.

Motivated by these findings, this study aims to investigate this generalisation problem using two intrusion detection datasets: UNSW-NB15 and TON\_IoT. Three supervised learning models – Random Forest, Logistic Regression, and Naive Bayes – are evaluated under both same-dataset and cross-dataset settings. The purpose is to compare model performance within individual datasets and examine whether that performance remains consistent across different datasets. By systematically comparing multiple machine learning models using standard evaluation metrics, this paper seeks to contribute to the development of more flexible, accurate, and reliable intrusion detection systems.

\section{Methods}

\begin{figure*}[t]
  \centering
  \includegraphics[width=\textwidth]{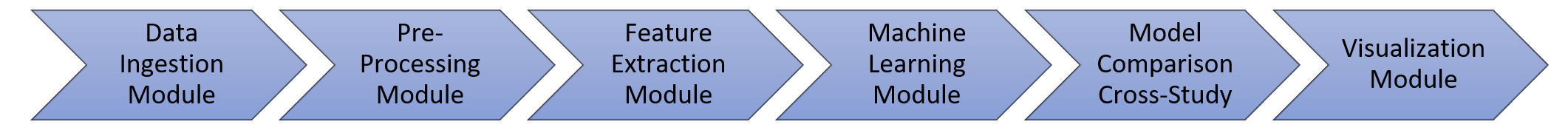}
  \caption{Development model used in this study.}
  \label{fig:workflow}
\end{figure*}

This study evaluates the generalisation capability of machine-learning-based intrusion detection models. The study is mainly focused on forecasting different types of network attacks, followed by a methodical process that included the collection of different datasets, preparing them, and creating and testing models. Figure \ref{fig:workflow} shows the workflow of our model.  Below is the description of the working procedure followed during the study.

\subsection{Datasets}

This study uses a convenience sampling method to collect datasets for model training and testing. Two publicly available datasets - UNSW-NB15 \cite{moustafa2015unsw}, and TON\_IoT \cite{moustafa2021new} are used due to their accessibility and relevance to intrusion detection. These datasets contain labeled network traffic with multiple attack types. 

The UNSW-NB15 dataset was used as one of the main benchmark datasets for intrusion detection. After preprocessing and cleaning, the dataset contained 257,673 records. The TON\_IoT dataset was used as the second benchmark dataset, representing IoT/IIoT network traffic. After duplicate removal and cleaning, the TON\_IoT dataset contained 190,474 records.

For same-dataset experiments, each dataset was processed using its full available cleaned feature set. For cross-dataset experiments, only the common features available in both datasets were retained. This was necessary because UNSW-NB15 and TON\_IoT do not share the same complete feature structure. The common feature set used for cross-dataset evaluation consisted of seven input features: duration, proto, service, src\_bytes, dst\_bytes, src\_pkts, and dst\_pkts. The label column was used as the target variable.

In this study, two evaluation settings were used: same-dataset evaluation and cross-dataset evaluation. In the same-dataset setting, each dataset was divided into training and testing subsets using a 70/30 split, where 70\% of the data was used for training and 30\% was used for testing. In the cross-dataset setting, models trained on one dataset were tested on the other dataset to examine generalisation across different network environments. Therefore, UNSW-NB15 and TON\_IoT were used both as benchmark datasets and as cross-evaluation datasets.

\subsection{Implementation and Tools}

All experiments were implemented in Python using standard machine learning libraries, including Pandas, NumPy, Scikit-learn, and Matplotlib/Seaborn. The development workflow used Jupyter Notebook, VS Code, and Python scripts executed in the experimental environment. These tools were used for data preprocessing, model training, evaluation, and result visualization.

\subsection{Procedure}

The procedure was executed in various phases including data preprocessing, cleaning, model selection, training, architectural design, and evaluation. Some of the important phases of the data processing are described below.

\subsubsection{Data preprocessing}
As an initial step, categorical variables (e.g., protocol type and attack labels) were converted into numerical form using label encoding. Numerical features were normalized using Min–Max normalization to reduce the effect of different feature scales. In addition, feature correlation analysis was conducted to examine inter-feature relationships and identify possible redundancy in the dataset. For the cross-dataset experiment, only the common feature columns shared by UNSW-NB15 and TON\_IoT were retained, being 7 out of aaa and bbb features, respectively.

\subsubsection{Data cleaning}
To improve data reliability, Infinite values (e.g., due to division-by-zero during feature generation) were first replaced with NaN. Rows containing missing values were then removed, and duplicate records were dropped from the datasets. This cleaning process was applied before model training to ensure that the models were trained only on complete and non-duplicated records. 

\subsubsection{Model selection and training}
For model comparison across the datasets, three supervised machine learning classifiers were used: Random Forest, Logistic Regression, and Naive Bayes. These models were selected to compare ensemble-based, linear, and probabilistic learning approaches. In the same-dataset experiments, each dataset was divided into training and testing subsets for model evaluation. In the cross-dataset experiments, models were trained on one dataset and tested directly on the other dataset using the aligned common feature set. Model performance was evaluated using accuracy, precision, recall, and F1-score.

\section{Results \& Discussion}

\subsection{Feature Correlation Analysis}

\begin{figure}[t]
\centering
\includegraphics[width=0.75\textwidth]{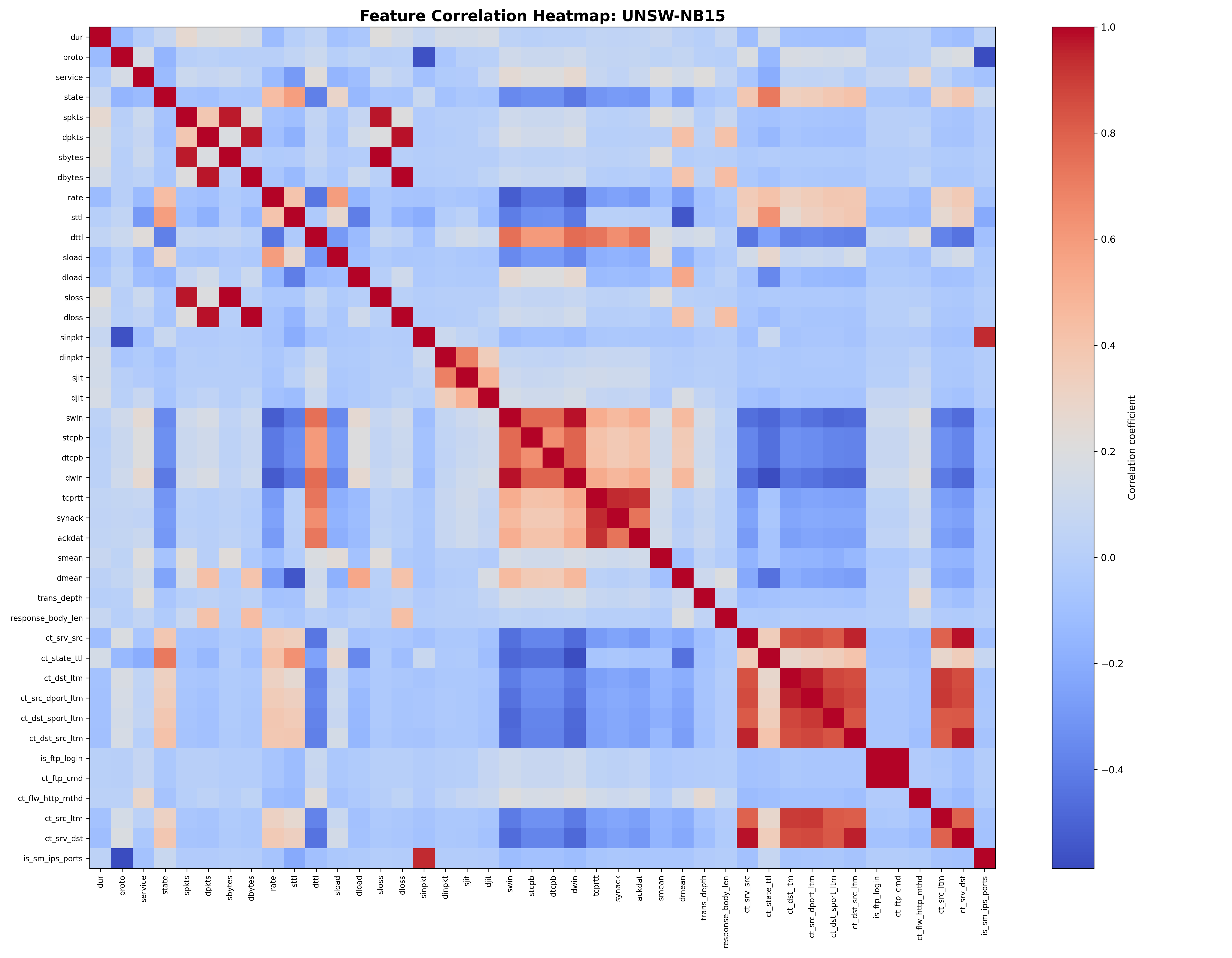}
\caption{Feature correlation heatmap of the UNSW-NB15 dataset.}
\label{fig:unsw-correlation}
\end{figure}

Before evaluating the models, feature-level relationships were examined to understand the structure of the dataset and identify possible redundancy among the input variables. Fig. \ref{fig:unsw-correlation} presents the correlation heatmap for the UNSW-NB15 dataset. The UNSW-NB15 correlation heatmap was used as a sample feature-level analysis to examine relationships among input variables and identify possible redundancy. The heatmap shows that several traffic-volume and flow-based features, in particular, byte-count and packet-count features such as sbytes, dbytes, spkts, and dpkts show noticeable positive correlations, indicating that they represent overlapping aspects of traffic volume.

A similar relationship can also be observed among statistical traffic features such as smean, dmean, sload, and dload, which describe the intensity and direction of network communication. Connection-tracking attributes, including features such as ct\_src\_ltm, ct\_dst\_ltm, and ct\_dst\_src\_ltm, also form correlated groups. These patterns suggest that the dataset contains partially redundant information, which is common in network-flow-based intrusion detection datasets.

However, the heatmap does not show extreme correlation across the entire feature set. Therefore, the features still provide useful and diverse information for model training. At the same time, the presence of correlated feature groups indicates that model performance may depend on dataset-specific feature relationships. These feature relationships provide useful context for understanding how the models use traffic-related information during classification.

\subsection{Same Dataset Evaluation}

In the same-dataset setting, each model was trained and tested on the same dataset. Table \ref{tab:unsw-metrics} presents the performance of the three classifiers on UNSW-NB15. Random Forest achieved the strongest overall performance, with 95.08\% accuracy, 96.21\% precision, 96.08\% recall, and 96.15\% F1-score. Logistic Regression also performed well, achieving 89.53\% accuracy and the highest recall among the three models at 97.22\%. Naive Bayes achieved the lowest accuracy at 81.47\%, but its precision and F1-score indicate that it still provided a reasonable baseline performance on this dataset.

\begin{table}[h]
\centering
\caption{Model performance on UNSW-NB15.}
\label{tab:unsw-metrics}
\begin{tabular}{lcccc}
\hline
Model & Accuracy & F1 & Precision & Recall \\
\hline
Random Forest & 0.9508 & 0.9614 & 0.9621 & 0.9607 \\
Logistic Regression & 0.8952 & 0.9222 & 0.8771 & 0.9722 \\
Na\"ive Bayes & 0.8146 & 0.8506 & 0.8761 & 0.8266 \\
\hline
\end{tabular}
\end{table}

\begin{table}[H]
\centering
\caption{Model performance on TON\_IoT.}
\label{tab:ton_iot-metrics}
\setlength{\tabcolsep}{6pt}
\renewcommand{\arraystretch}{0.95}
\begin{tabular}{lcccc}
\hline
Model & Accuracy & F1 & Precision & Recall \\
\hline
Random Forest & 0.9979 & 0.9987 & 0.9981 & 0.9992 \\
Logistic Regression & 0.8808 & 0.9274 & 0.8662 & 0.9978 \\
Na\"ive Bayes & 0.7731 & 0.8705 & 0.7711 & 0.9992 \\
\hline
\end{tabular}
\end{table}

Table \ref{tab:ton_iot-metrics}  shows the same-dataset results on the TON\_IoT dataset. Random Forest again produced the best overall performance, achieving 99.79\% accuracy, 99.82\% precision, 99.92\% recall, and 99.87\% F1-score. Logistic Regression achieved 88.08\% accuracy and a high recall of 99.78\%, showing that it detected most positive cases but with lower precision than Random Forest. Naive Bayes achieved 77.31\% accuracy, which was the lowest among the three models, although its recall was also very high at 99.92\%. This indicates that both Logistic Regression and Naive Bayes were able to detect many attack samples, but Random Forest provided the most balanced and accurate result.

\subsection{Feature Importance Analysis}

\begin{figure}[htbp]
\centering
\includegraphics[width=\columnwidth]{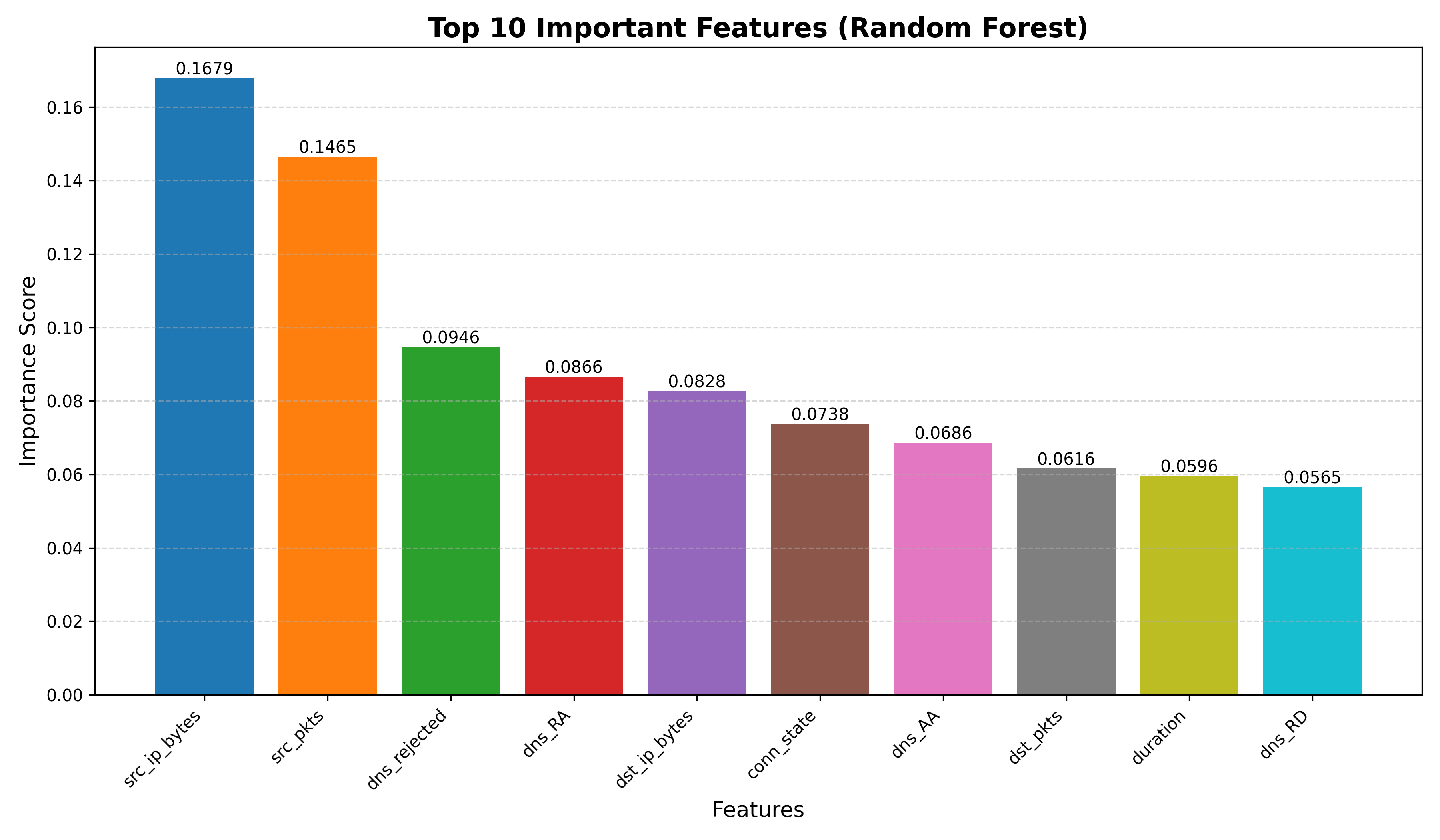}
\caption{Top ten important features identified by the Random Forest model on the TON\_IoT dataset.}
\label{fig:ton-feature-importance}
\end{figure}

Since Random Forest performed the best across both datasets, feature importance analysis was conducted using the trained Random Forest model on TON\_IoT as a representative model-interpretation example. Feature importance analysis helps identifying which features contributed most strongly to the classification decision. Fig. \ref{fig:ton-feature-importance} shows the top-ranked features from the TON\_IoT dataset. The result indicates that traffic-volume-related features were among the most influential variables. In particular, src\_ip\_bytes and src\_pkts received the highest importance scores, showing that source-side byte volume and packet activity played a major role in intrusion detection.

DNS-related features also had strong influence on the model. Features such as dns\_rejected, dns\_RA, and dns\_AA appeared among the top contributors, suggesting that DNS behaviour was useful for separating normal and malicious traffic in the TON\_IoT dataset. In addition, connection-level and direction-based traffic features, including conn\_state, dst\_ip\_bytes, and dst\_pkts, contributed to the model’s decision process. This result helps explain the high same-dataset performance of Random Forest on TON\_IoT. 

\subsection{Comparative Analysis}

\begin{table}[!t]
\centering
\caption{Comparison of existing studies with the proposed approach.}
\label{tab:comparison}
\setlength{\tabcolsep}{4pt}
\renewcommand{\arraystretch}{0.95}

\begin{tabularx}{\textwidth}{@{}
>{\raggedright\arraybackslash}X
>{\raggedright\arraybackslash}p{1.9cm}
>{\centering\arraybackslash}p{1.25cm}
>{\centering\arraybackslash}p{1.65cm}
@{}}

\toprule
Existing work & Dataset & ML model & Accuracy (\%) \\
\midrule

Lopez-Martin et al. (2019)~\cite{lopez2019shallow}
  & UNSW-NB15 & RF & 74.00 \\
\cmidrule(lr){1-4}

Zhou et al. (2018)~\cite{zhou2018deep}
  & UNSW-NB15 & LR & 89.47 \\
\cmidrule(lr){1-4}

Moustafa \& Slay (2016)~\cite{moustafa2016evaluation}
  & UNSW-NB15 & LR & 83.15 \\
\cmidrule(lr){1-4}

Janarthanan \& Zargari (2017)~\cite{janarthanan2017feature}
  & UNSW-NB15 & RF & 75.66 \\
\cmidrule(lr){1-4}

Ajagbe et al. (2024)~\cite{ajagbe2024intrusion}
  & UNSW-NB15 & RF & 90.39 \\
  &            & LR & 69.41 \\
\cmidrule(lr){1-4}

Azeroual et al. (2022)~\cite{azeroual2022analysis}
  & UNSW-NB15 & LR & 86.44 \\
  &            & NB & 77.33 \\
\cmidrule(lr){1-4}

Dharini et al. (2026)~\cite{dharini2026efficient}
  & TON\_IoT & RF & 98.66 \\
\cmidrule(lr){1-4}

Almotairi et al. (2024)~\cite{almotairi2024enhancing}
  & TON\_IoT & RF & 98.20 \\
  &          & NB & 80.46 \\
\cmidrule(lr){1-4}

Gad et al. (2021)~\cite{gad2021intrusion}
  & TON\_IoT & RF & 97.90 \\
  &          & LR & 86.30 \\
  &          & NB & 46.40 \\

\midrule
Our approach
  & UNSW-NB15 & RF & 95.08 \\
  &           & LR & 89.53 \\
  &           & NB & 81.47 \\
\cmidrule(lr){2-4}
  & TON\_IoT  & RF & 99.79 \\
  &           & LR & 88.08 \\
  &           & NB & 77.31 \\

\bottomrule
\end{tabularx}
\end{table}

Table~\ref{tab:comparison} compares the same-dataset accuracy obtained in this study with selected previous studies on UNSW-NB15 and TON\_IoT. For UNSW-NB15, the Random Forest achieved 95.08\% accuracy, which is the highest accuracy achieved amongst the compared works using the Random Forest model, and overall.. Logistic Regression and Na\"ive Bayes achieved 89.53\% and 81.47\% accuracies respectively, which are also the highest accuracies among the compared works using those models.

Random Forest performed the best for TON\_IoT as well. The model achieved 99.79\% accuracy, which is higher than the results reported by Dharini et al. \cite{dharini2026efficient} (98.66\%), Almotairi et al. \cite{almotairi2024enhancing} (98.20\%), and Gad et al. \cite{gad2021intrusion} (97.90\%). Logistic Regression achieved 88.08\%, which is also higher than the 86.30\% reported by Gad et al. \cite{gad2021intrusion}. Naive Bayes achieved 77.31\%, which is higher than the 46.40\% reported by Gad et al. \cite{gad2021intrusion}, but lower than the 80.46\% reported by Almotairi et al. \cite{almotairi2024enhancing}.

Overall, the comparison shows that our model exhibited superior performance in most cases in the same-dataset settings, and the results of this study are competitive with previous research works. Random Forest remains the strongest model in both datasets, while Logistic Regression also performs well, especially on UNSW-NB15 and TON\_IoT when compared with selected previous studies.

\subsection{Cross-Dataset Evaluation}

In order to further test the capability of the models, we performed a cross-dataset evaluation to examine whether the models can generalize beyond the dataset used for training. In the cross-dataset setting, the models were trained on one dataset and tested on the other dataset. Since UNSW-NB15 and TON\_IoT have different feature structures, only the common features available in both datasets were used for this evaluation. The common features were duration, proto, service, src\_bytes, dst\_bytes, src\_pkts, and dst\_pkts.

\begin{table}[h]
\centering
\caption{Cross-dataset performance: UNSW-NB15 to TON\_IoT.}
\label{tab:unsw-to-ton-metrics}
\begin{tabular}{lcccc}
\hline
Model & Accuracy & F1 & Precision & Recall \\
\hline
Random Forest & 0.3807 & 0.4898 & 0.6841 & 0.3814 \\
Logistic Regression & 0.7779 & 0.8751 & 0.7790 & 0.9982 \\
Na\"ive Bayes & 0.2585 & 0.1020 & 0.9066 & 0.0541 \\
\hline
\end{tabular}
\end{table}

Table \ref{tab:unsw-to-ton-metrics} presents the results when the models were trained on UNSW-NB15 and tested on TON\_IoT. Logistic Regression achieved the best performance in this direction, with 77.79\% accuracy, 77.90\% precision, 99.82\% recall, and 87.51\% F1-score. Random Forest, despite being the strongest model in same-dataset testing, dropped to 38.07\% accuracy and 48.98\% F1-score. Naive Bayes showed the weakest overall result, with 25.85\% accuracy and only 10.20\% F1-score. Although Naive Bayes had high precision, its recall was only 5.41\%, indicating that it failed to detect most positive cases in the TON\_IoT test data.

\begin{table}[h]
\centering
\caption{Cross-dataset performance: TON\_IoT to UNSW-NB15.}
\label{tab:ton-to-unsw-metrics}
\begin{tabular}{lcccc}
\hline
Model & Accuracy & F1 & Precision & Recall \\
\hline
Random Forest & 0.3954 & 0.5268 & 0.5270 & 0.5267 \\
Logistic Regression & 0.3824 & 0.4902 & 0.5188 & 0.4646 \\
Na\"ive Bayes & 0.3658 & 0.0573 & 0.5726 & 0.0302 \\
\hline
\end{tabular}
\end{table}

Table \ref{tab:ton-to-unsw-metrics} presents the results when the models were trained on TON\_IoT and tested on UNSW-NB15. In this setup, all three models showed very weak performance. Random Forest achieved the highest accuracy at 39.54\%, with precision, recall, and F1-score all close to 52\%. Logistic Regression achieved 38.24\% accuracy and 49.02\% F1-score. Naive Bayes achieved 36.58\% accuracy, but its recall and F1-score were very low, at 3.02\% and 5.73\%, respectively. This shows that Naive Bayes performed poorly in identifying positive cases when tested on UNSW-NB15 after training on TON\_IoT.

The cross-dataset performance drop can be partly explained by the differences between the two datasets. First, the cross-dataset experiments used only seven common features shared by UNSW-NB15 and TON\_IoT. This reduced feature set may have removed dataset-specific information that was useful in the same-dataset experiments. As a result, the models had less discriminative information available during cross-dataset testing than during same-dataset testing.

Second, the two datasets differ in traffic generation environment, feature distribution, and attack behaviour. The correlation analysis showed strong relationships among some traffic-volume and connection-related features in UNSW-NB15, while the Random Forest feature-importance result showed high dependence on traffic-volume, packet-count, DNS, and connection-state features in TON\_IoT. In addition, the packet-count distribution comparison shows a clear difference between the two datasets, with UNSW-NB15 having a wider and more dispersed distribution and TON\_IoT being more concentrated around lower values.

These differences suggest that the models learned patterns that were useful within one dataset but less reliable when applied to another dataset. This is especially visible for Random Forest, which achieved the best same-dataset performance but dropped sharply in both cross-dataset directions. Therefore, the results indicate that high same-dataset accuracy does not necessarily reflect strong generalisation when the testing data comes from a different network environment.

\section{Conclusion}

In this study, we evaluated machine-learning-based intrusion detection using two labelled intrusion datasets, UNSW-NB15 and TON\_IoT. The study used supervised learning models, namely Random Forest, Logistic Regression, and Naive Bayes, for evaluation under both same-dataset and cross-dataset settings. In the same-dataset experiments, Random Forest produced the strongest overall results, achieving near perfect accuracies on UNSW-NB15 and on TON\_IoT when trained seperately.

Comparatively, the cross-dataset evaluation showed that model performance changed significantly when the training and testing datasets were different. Despite achieving the best accuracies in the same-dataset experiments, Random Forest's performance plummeted to under 40\% when trained on one and tested on the other.

The key contribution is therefore not another comparison of classifiers, but the exposure of a noteable generalisation problem in cybersecurity machine learning. This has direct relevance to affective computing and human-centric AI. In affective computing, models must infer latent states from noisy, context-dependent behavioural, visual, physiological, and temporal signals; intrusion detection faces an analogous problem in identifying malicious activity from noisy, shifting, and environment-dependent network behaviour. Techniques developed for behavioural signal analysis, anomaly detection, domain adaptation, uncertainty-aware modelling, and robust cross-context evaluation can therefore be leveraged to improve cyber-security systems. This connection is particularly important for human-centred information security, where cyber risk often emerges from interactions between users, devices, networks, and adaptive adversaries.

Some challenges that might have contributed to the poor performance of the cross-dataset experiment were differences in dataset structure, feature distribution, attack behaviour, and the limited number of common features available between UNSW-NB15 and TON\_IoT. Since the cross-dataset evaluation used only the shared feature set, some dataset-specific information used in the same-dataset experiments may not have been available during cross-dataset testing. Despite these limitations, the results demonstrate that model choice, dataset characteristics, and feature representation strongly influence IDS performance. Therefore, consistent preprocessing and cross-dataset evaluation are essential for meaningful comparison of real-world viability. Future work must focus on improving cross-dataset generalisation through better feature alignment, feature selection, domain adaptation, and more robust learning approaches to address evolving information security threats.

Ultimately, this study reveals a critical "generalization gap" in machine-learning-based intrusion detection that poses a significant challenge to real-world network defense. While models like Random Forest can achieve near-perfect accuracy within a single, controlled environment, the sharp decline in performance during cross-dataset evaluations proves that high benchmark scores often mask a failure to adapt to diverse network traffic. For the information security community to build truly resilient systems, it is no longer enough to optimize for static datasets; the focus must shift toward developing adaptive models that can generalize across evolving IoT and enterprise environments. This includes improved feature alignment, domain adaptation, uncertainty modelling, and multimodal extensions that combine network-level signals with user, device, and behavioural context. By framing intrusion detection as a generalisable behavioural
modelling problem, this study provides a foundation for stronger collaboration between affective computing and information security research.

\section*{Declaration of Generative AI and AI-Assisted Technologies in the Manuscript Preparation Process}

During the preparation of this manuscript, the authors used ChatGPT as a language-support tool to assist with grammar checking, sentence restructuring, and clarity improvement. The tool was not used to generate any experimental results. After using this tool, the authors carefully reviewed, edited, and validated the content for technical accuracy, originality, and consistency with the reported experiments. The authors take full responsibility for the content of the manuscript.
%
%
%
\bibliographystyle{splncs04}
%
\bibliography{Bibliography}

@article{aljanabi2021intrusion,
  title={Intrusion detection: A review},
  author={Aljanabi, Mohammad and Ismail, Mohd Arfian and Hasan, Raed Abdulkareem and Sulaiman, Junaida},
  journal={Mesopotamian Journal of CyberSecurity},
  volume={2021},
  pages={1--4},
  year={2021}
}

@article{kenyon2020public,
  title={Are public intrusion datasets fit for purpose characterising the state of the art in intrusion event datasets},
  author={Kenyon, Anthony and Deka, Lipika and Elizondo, David},
  journal={Computers \& Security},
  volume={99},
  pages={102022},
  year={2020},
  publisher={Elsevier}
}

@article{saranya2020performance,
  title={Performance analysis of machine learning algorithms in intrusion detection system: A review},
  author={Saranya, T and Sridevi, S and Deisy, C and Chung, Tran Duc and Khan, MKA Ahamed},
  journal={Procedia Computer Science},
  volume={171},
  pages={1251--1260},
  year={2020},
  publisher={Elsevier}
}

@article{gumucsbacs2020comprehensive,
  title={A comprehensive survey of databases and deep learning methods for cybersecurity and intrusion detection systems},
  author={G{\"u}m{\"u}{\c{s}}ba{\c{s}}, Dilara and Y{\i}ld{\i}r{\i}m, Tulay and Genovese, Angelo and Scotti, Fabio},
  journal={IEEE Systems Journal},
  volume={15},
  number={2},
  pages={1717--1731},
  year={2020},
  publisher={IEEE}
}

@article{khan2024ai,
  title={AI-Driven Threat Detection: A Brief Overview of AI Techniques in Cybersecurity},
  author={Khan, Muhammad Ismaeel and Arif, Aftab and Khan, Ali Raza A},
  journal={BIN: Bulletin of Informatics},
  volume={2},
  number={2},
  pages={248--61},
  year={2024}
}

@inproceedings{adek2020survey,
  title={A survey on the accuracy of machine learning techniques for intrusion and anomaly detection on public data sets},
  author={Adek, Rizal Tjut and Ula, Munirul},
  booktitle={2020 International Conference on Data Science, Artificial Intelligence, and Business Analytics (DATABIA)},
  pages={19--27},
  year={2020},
  organization={IEEE}
}

@article{maseer2021benchmarking,
  title={Benchmarking of machine learning for anomaly based intrusion detection systems in the CICIDS2017 dataset},
  author={Maseer, Ziadoon Kamil and Yusof, Robiah and Bahaman, Nazrulazhar and Mostafa, Salama A and Foozy, Cik Feresa Mohd},
  journal={IEEE access},
  volume={9},
  pages={22351--22370},
  year={2021},
  publisher={IEEE}
}

@inproceedings{panwar2022intrusion,
  title={An intrusion detection model for cicids-2017 dataset using machine learning algorithms},
  author={Panwar, Shailesh Singh and Raiwani, YP and Panwar, Lokesh Singh},
  booktitle={2022 International Conference on Advances in Computing, Communication and Materials (ICACCM)},
  pages={1--10},
  year={2022},
  organization={IEEE}
}

@inproceedings{almseidin2017evaluation,
  title={Evaluation of machine learning algorithms for intrusion detection system},
  author={Almseidin, Mohammad and Alzubi, Maen and Kovacs, Szilveszter and Alkasassbeh, Mouhammd},
  booktitle={2017 IEEE 15th international symposium on intelligent systems and informatics (SISY)},
  pages={000277--000282},
  year={2017},
  organization={IEEE}
}

@article{kanimozhi2019calibration,
  title={Calibration of various optimized machine learning classifiers in network intrusion detection system on the realistic cyber dataset CSE-CIC-IDS2018 using cloud computing},
  author={Kanimozhi, V and Jacob, T Prem},
  journal={International Journal of Engineering Applied Sciences and Technology},
  volume={4},
  number={6},
  pages={2455--2143},
  year={2019},
  publisher={IJEAST}
}

@misc{IBM, url={https://www.ibm.com/reports/data-breach}, journal={IBM}}

@article{lu2025adversarial,
  title={Adversarial attacks based on time-series features for traffic detection},
  author={Lu, Hongyu and Liu, Jiajia and Peng, Jimin and Lu, Jiazhong},
  journal={Computers \& Security},
  volume={148},
  pages={104175},
  year={2025},
  publisher={Elsevier}
}

@article{axelsson2000intrusion,
  title={Intrusion detection systems: A survey and taxonomy},
  author={Axelsson, Stefan},
  year={2000},
  publisher={Technical Report 99--15, Chalmers Univ}
}

@inproceedings{moustafa2015unsw,
  title={UNSW-NB15: a comprehensive data set for network intrusion detection systems (UNSW-NB15 network data set)},
  author={Moustafa, Nour and Slay, Jill},
  booktitle={2015 military communications and information systems conference (MilCIS)},
  pages={1--6},
  year={2015},
  organization={Ieee}
}

@misc{moustafa2021new,
  title={A new distributed architecture for evaluating AI-based security systems at the edge: Network TON\_IoT datasets. Sustain. Cities Soc. 72, 102994 (2021)},
  author={Moustafa, N},
  year={2021}
}

@article{almotairi2024enhancing,
  title={Enhancing intrusion detection in IoT networks using machine learning-based feature selection and ensemble models},
  author={Almotairi, Ayoob and Atawneh, Samer and Khashan, Osama A and Khafajah, Nour M},
  journal={Systems Science \& Control Engineering},
  volume={12},
  number={1},
  pages={2321381},
  year={2024},
  publisher={Taylor \& Francis}
}

@article{moustafa2016evaluation,
  title={The evaluation of Network Anomaly Detection Systems: Statistical analysis of the UNSW-NB15 data set and the comparison with the KDD99 data set},
  author={Moustafa, Nour and Slay, Jill},
  journal={Information Security Journal: A Global Perspective},
  volume={25},
  number={1-3},
  pages={18--31},
  year={2016},
  publisher={Taylor \& Francis}
}

@inproceedings{janarthanan2017feature,
  title={Feature selection in UNSW-NB15 and KDDCUP'99 datasets},
  author={Janarthanan, Tharmini and Zargari, Shahrzad},
  booktitle={2017 IEEE 26th international symposium on industrial electronics (ISIE)},
  pages={1881--1886},
  year={2017},
  organization={IEEE}
}

@article{ajagbe2024intrusion,
  title={Intrusion detection: a comparison study of machine learning models using unbalanced dataset},
  author={Ajagbe, Sunday Adeola and Awotunde, Joseph Bamidele and Florez, Hector},
  journal={SN Computer Science},
  volume={5},
  number={8},
  pages={1028},
  year={2024},
  publisher={Springer}
}

@article{lopez2019shallow,
  title={Shallow neural network with kernel approximation for prediction problems in highly demanding data networks},
  author={Lopez-Martin, Manuel and Carro, Belen and Sanchez-Esguevillas, Antonio and Lloret, Jaime},
  journal={Expert Systems with Applications},
  volume={124},
  pages={196--208},
  year={2019},
  publisher={Elsevier}
}

@inproceedings{zhou2018deep,
  title={Deep learning approach for cyberattack detection},
  author={Zhou, Yiyun and Han, Meng and Liu, Liyuan and He, Jing Selena and Wang, Yan},
  booktitle={IEEE INFOCOM 2018-IEEE conference on computer communications workshops (INFOCOM WKSHPS)},
  pages={262--267},
  year={2018},
  organization={IEEE}
}

@inproceedings{azeroual2022analysis,
  title={Analysis of UNSW-NB15 Datasets Using Machine Learning Algorithms},
  author={Azeroual, Hakim and Belghiti, Imane Daha and Berbiche, Naoual},
  booktitle={International Conference on Digital Technologies and Applications},
  pages={199--209},
  year={2022},
  organization={Springer}
}

@article{dharini2026efficient,
  title={Efficient detection of intrusions in TON-IoT dataset using hybrid feature selection approach},
  author={Dharini, N and Janani, VS and Katiravan, Jeevaa},
  journal={Scientific Reports},
  year={2026},
  publisher={Nature Publishing Group UK London}
}

@article{gad2021intrusion,
  title={Intrusion detection system using machine learning for vehicular ad hoc networks based on ToN-IoT dataset},
  author={Gad, Abdallah R and Nashat, Ahmed A and Barkat, Tamer M},
  journal={IEEE access},
  volume={9},
  pages={142206--142217},
  year={2021},
  publisher={IEEE}
}

@inproceedings{li2017intrusion,
  title={Intrusion detection using convolutional neural networks for representation learning},
  author={Li, Zhipeng and Qin, Zheng and Huang, Kai and Yang, Xiao and Ye, Shuxiong},
  booktitle={International conference on neural information processing},
  pages={858--866},
  year={2017},
  organization={Springer}
}
\end{document}